\documentclass[reprint,prb,amsmath,amssymb,superscriptaddress]{revtex4-2}
\usepackage{graphicx}
\usepackage{bm}
\usepackage[utf8]{inputenc}

\usepackage[hidelinks]{hyperref}
\usepackage{orcidlink}

\newcommand{\bb}{\mathbf{b}}
\newcommand{\KK}{\mathbf{K}}

\begin{document}

\title{Torsion, valley holonomy, and Berry phases of quasiparticles\\ encircling dislocations and defect complexes in graphene}
\author{G. Q. Garcia\,\orcidlink{0000-0003-3562-0317}}
\email{gqgarcia99@gmail.com}
\affiliation{Centro de Ci\^encias, Tecnologia e Sa\'ude, Universidade Estadual da Para\'iba, 58233-000, Araruna, PB, Brazil.}
\author{A. M. de M. Carvalho\,\orcidlink{0009-0006-3540-0364}}
\email{alexandre@fis.ufal.br}
\affiliation{Instituto de F\'{\i}sica, Universidade Federal de Alagoas, 57072-970, Macei\'o, AL, Brazil.}
\author{C. Furtado\,\orcidlink{0000-0002-3455-4285}}
\email{furtado@fisica.ufpb.br}
\affiliation{Departamento de F\'isica, Universidade Federal da Para\'iba, Caixa Postal 5008, 58051-900, Jo\~ao Pessoa, PB, Brazil}

\date{\today}

\begin{abstract}
Real graphene sheets contain pentagon--heptagon pairs, Stone--Wales defects, divacancies, and square--octagon structures. We ask what geometric phase a low-energy quasiparticle picks up when it goes around one of these defects, and we restrict attention throughout to loops that enclose the defect. Following Anandan, the Berry phase in a space with torsion is the holonomy of the Poincar\'e connection $\theta^aP_a+\frac12\omega^{ab}M_{ab}$. Its translational part couples momentum to torsion, so a loop that encloses a torsion flux acquires the phase $P_a\int T^a$. In the Katanaev--Volovich theory of defects, a dislocation is a flat space whose torsion sits entirely in the core, and for graphene the momentum that couples to it is the crystal momentum $\hbar(\tau_3\KK+\mathbf q)$. We find that the Anandan phase of an isolated dislocation never vanishes. It contains a topological valley phase $e^{\mp2\pi i/3}$, present whenever the Burgers vector $\bb$ lies outside the $\sqrt3\times\sqrt3$ superlattice, and a dispersive phase $e^{-i\mathbf q\cdot\bb}$, present for every $\bb\neq0$. The topological part is the lattice result of Mesaros \textit{et al.}, which here appears as a torsion-flux phase; dislocations with $\bb$ in the superlattice, such as 4--8 cores, keep only the dispersive part. Composing the Volterra elements of the non-hexagonal rings extends the result to arbitrary defect complexes. When the net Frank angle is zero the holonomy is again $\exp(i\tau_3\KK\cdot\bb)$, now with the net Burgers vector, and the valley flux obeys a three-coloring rule: pentagons and heptagons contribute through the Kekul\'e color of the sites they occupy, while squares and octagons never contribute. Atomic models confirm the rule and show that the ring content alone does not fix the phase. A 5--8--5 core, for example, is invisible as a divacancy and as a dislocation with $|\bb|=\sqrt3a$, but carries a flux of $\pm1/3$ as a dislocation with $\bb=2\mathbf a_1$. Finally, the torsion trace produces no geometric phase. A vector coupling to it would break time reversal, and the axial coupling that symmetry allows only causes anisotropic short-range scattering.
\end{abstract}

\maketitle

\section{Introduction}

Close to the two inequivalent corners $\pm\KK$ of the Brillouin zone, electrons in graphene obey a massless Dirac equation in $2+1$ dimensions~\cite{voz6,nov,castro,geim}. This has made graphene a natural setting for studying how lattice defects act as geometry. Disclinations (pentagons and heptagons) bring curvature together with non-Abelian gauge fields that rotate the valley and sublattice indices~\cite{lamert,def1,def2,sitenko}. Smooth deformations generate pseudo-gauge fields that enter the two valleys with opposite signs~\cite{moro,mopu,vozrep,dejuan2012}; in the language of the Dirac equation they are axial gauge fields~\cite{jackiw,cortijo2015}. Graphene in curved and defected geometries has accordingly received a great deal of attention~\cite{claudio1,bf19,bueno,bf30,bent,b15,sharapov}.

Because their energy is finite, dislocations are present in every real crystal; in graphene they have been imaged directly as pentagon--heptagon (5--7) pairs~\cite{warner,tor}. In the geometric theory of defects of Katanaev and Volovich (KV)~\cite{kat,katrev,kleinert,moraesG2}, a dislocation is described by a flat space with torsion, and its Burgers vector is the flux of the torsion two-form. Working on the lattice, Mesaros, Sadri, and Zaanen showed that a dislocation gives graphene electrons a Berry phase of $0$ or $\pm1/3$ of a flux quantum, with opposite signs in the two valleys~\cite{mes2}, and that this ``valley holonomy'' is the translational part of parallel transport in a Riemann--Cartan space~\cite{mes1}. Torsional and dislocation responses have since become important in the study of topological phases~\cite{ran,juricic,hughes}, and wave-front dislocations in Friedel oscillations have been used to measure the Berry phase of graphene directly~\cite{dutreix}. Real samples, however, contain much more than isolated 5--7 pairs. Stone--Wales (5--7--7--5) defects, reconstructed divacancies (5--8--5), line defects built from pentagons and octagons, and grain boundaries made of 5--7 and 4--8 units are all observed~\cite{banhart,kotakoski,lahiri,warner}. A quasiparticle that travels around any of these structures probes its holonomy, and the question we address is which of them leave a topological phase behind, and how large that phase is.

A separate line of work couples fermions to torsion through nonminimal terms~\cite{shap}. The minimal Dirac coupling sees only the totally antisymmetric (axial) part of the torsion, which vanishes for an edge dislocation, and in Ref.~\cite{bf31} a nonminimal coupling to the torsion trace was used to obtain an Abelian geometric phase for a neutral particle in the background of a spiral dislocation, a distortion of a circle into a spiral~\cite{valanis}. The geometry of that background, which differs from the one considered here, is discussed in the Supplemental Material~\cite{SM}.

How a quantum phase should be defined in the presence of torsion was worked out by Anandan~\cite{anandan77,anandan94,anandan96}. Along a path $\gamma$ the wave function picks up the Poincar\'e-group element
\begin{equation}
\Phi_\gamma=\mathcal P\exp\left[-\frac{i}{\hbar}\int_\gamma\left(\theta^a{}_\mu P_a+\frac12\,\omega_\mu{}^{ab}M_{ab}\right)dx^\mu\right],
\label{eq:anandan}
\end{equation}
where $P_a$ and $M_{ab}$ generate translations and rotations. The curvature of this connection is $T^aP_a+\frac12R^{ab}M_{ab}$, so torsion plays the role of field strength for the translations. On a closed loop that encloses a torsion flux, a particle with momentum $P_a$ therefore acquires the topological phase $P_a\int T^a/\hbar$, a translational counterpart of the Aharonov--Bohm phase~\cite{anandan96}. This is the generalized Berry phase we use. We consider only loops $C$ that enclose the defect: loops that do not enclose it have trivial holonomy, and loops that cross the core fall outside the continuum description.

Here we apply Anandan's phase operator to graphene within the KV description, keeping track of the distributional character of the torsion. The main results are the following.
(i) \textit{The Anandan phase of an isolated dislocation never vanishes.} On any loop around the core it is $\exp(-i\tau_3\KK\cdot\bb)\exp(-i\mathbf q\cdot\bb)$. The topological factor, $e^{\mp2\pi i/3}$ in the two valleys, is nontrivial whenever $\bb$ lies outside the $\sqrt3\times\sqrt3$ superlattice, and so for every minimal 5--7 dislocation. It coincides with the lattice result~\cite{mes2,mes1}, which we now derive as a torsion-flux phase. The dispersive factor is nontrivial for every $\bb\neq0$, so even dislocations with $\bb$ in the superlattice, which carry no valley flux, have a nonzero Anandan phase. The core behaves as an Aharonov--Bohm flux tube with field strength $K_aT^a$, and its flux does not depend on how the core is built.
(ii) For any defect complex with zero net Frank angle, the holonomy is $\exp(i\tau_3\KK\cdot\bb_{\rm net})$, and the valley flux follows a three-coloring rule to which only pentagons and heptagons contribute.
(iii) Atomic models of seven defects (5--7, 4--8, 5--5--8, Stone--Wales, and the divacancy) obey the rule and show that the ring content does not determine the phase.
(iv) The torsion trace carries no flux. Coupling to it vectorially breaks time-reversal symmetry, and the axial coupling that remains describes short-range scattering, which we calculate.
The algebra is given in full in the Supplemental Material~\cite{SM}, together with a script that checks each step symbolically.

Section~\ref{sec:geometry} sets up the geometry of the edge dislocation. Section~\ref{sec:holonomy} derives the valley holonomy and shows that the core acts as a valley flux tube. Section~\ref{sec:complexes} turns to loops around general defect complexes and to the atomic models. Section~\ref{sec:nonminimal} discusses the nonminimal coupling: its symmetry, its vanishing geometric phase, and the scattering it produces. Section~\ref{sec:conclusions} summarizes.

\section{Geometry of an edge dislocation}\label{sec:geometry}

\begin{figure}[t]
\includegraphics[width=\columnwidth]{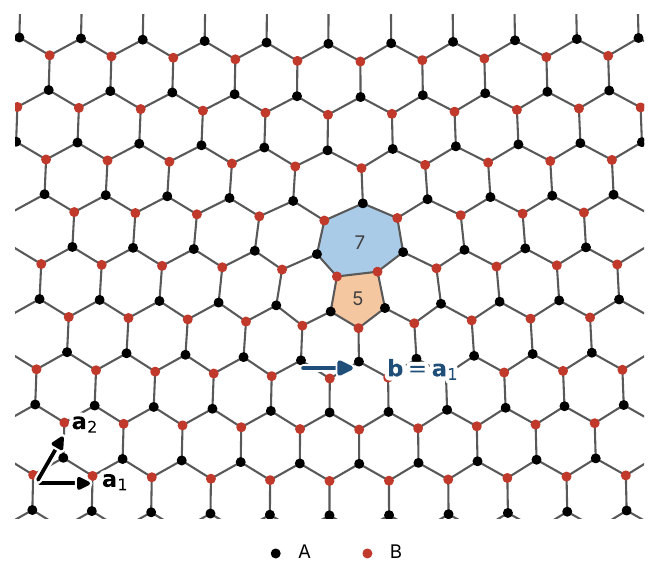}
\caption{Edge dislocation in graphene with Burgers vector $\bb=\mathbf a_1$ along $x$. One lattice period of material was removed in the upper half plane and the lattice was relaxed with the isotropic elastic field (Poisson ratio $0.17$). The core is a pentagon--heptagon pair. Black and red dots mark the $A$ and $B$ sublattices. The bond shared by the pentagon and the heptagon joins two $B$ sites, so the bipartite structure is broken only at the core.}
\label{fig:lattice}
\end{figure}

We set $\hbar=v_F=1$ and restore these constants when quoting numbers. An edge dislocation with Burgers vector $\bb=(b,0)$ in the plane of the sheet, with the defect line perpendicular to it, is described by the line element~\cite{puntigam,kat}
\begin{equation}
ds^2 = dt^2-\left(dx+\chi\,d\varphi\right)^2-dy^2,\qquad \chi=\frac{b}{2\pi},
\label{eq:metric}
\end{equation}
where $d\varphi=(x\,dy-y\,dx)/r^2$ and $r^2=x^2+y^2$. The signature is $(+,-,-)$, which matches the Dirac matrices of Sec.~\ref{sec:holonomy}. Expanding Eq.~(\ref{eq:metric}) gives the form used in Ref.~\cite{puntigam}. Figure~\ref{fig:lattice} shows the corresponding lattice.

A convenient frame is $\theta^0=dt$, $\theta^1=dx+\chi\,d\varphi=A\,dx+B\,dy$, and $\theta^2=dy$, with
\begin{equation}
A=1-\frac{\chi y}{r^2},\qquad B=\frac{\chi x}{r^2}.
\label{eq:AB}
\end{equation}
The dual frame is
\begin{equation}
e_1=\frac{1}{A}\,\partial_x,\qquad e_2=\partial_y-\frac{B}{A}\,\partial_x .
\label{eq:dual}
\end{equation}
In terms of the multivalued crystal coordinate $X=x+\chi\varphi$ and $Y=y$, the dual frame becomes $e_1=\partial_X$ and $e_2=\partial_Y$. Away from the core the space is flat, and the frame is just the Cartesian frame of the undeformed crystal. The determinant $\det\theta=A$ vanishes on a circle of radius $\chi/2=|\bb|/4\pi\approx 0.02$~nm that touches the origin. This lies well inside the core, where the continuum description does not hold in any case.

Because the curvature vanishes, we can choose the spin connection $\omega^a{}_b=0$ (the teleparallel, or Weitzenb\"ock, gauge). The Cartan structure equations then give the torsion two-form as $T^a=d\theta^a$. Away from the origin $d\theta^1=0$, whereas $\oint\theta^1=2\pi\chi=b$ on any loop around it. It follows that
\begin{equation}
T^1=b\,\delta^2(\mathbf r)\,dx\wedge dy=\frac{b}{2\pi}\,\delta(\rho)\,d\rho\wedge d\varphi ,
\label{eq:torsion}
\end{equation}
in Cartesian and polar coordinates, with $\int_0^\infty\delta(\rho)\,d\rho=1$. Note that a point source in two dimensions carries no angular delta function. In this language the Burgers vector is the flux of the torsion two-form.

In $D=2+1$ dimensions the torsion splits into a trace vector $T_\mu=T^{\beta}{}_{\mu\beta}$, an axial pseudoscalar $S=\epsilon^{\lambda\mu\nu}T_{\lambda\mu\nu}$, and a traceless remainder $q_{\lambda\mu\nu}$:
\begin{equation}
T_{\lambda\mu\nu}=\tfrac12\left(g_{\lambda\nu}T_\mu-g_{\lambda\mu}T_\nu\right)+\tfrac16\,\epsilon_{\lambda\mu\nu}S+q_{\lambda\mu\nu}.
\label{eq:decomp}
\end{equation}
The coefficients are not the familiar four-dimensional ones~\cite{shap}; the trace coefficient is $1/(D-1)$. For the torsion~(\ref{eq:torsion}) the only components are $T^1{}_{12}=-T^1{}_{21}=b\,\delta^2(\mathbf r)$ (to leading order at the core). Hence $S=0$, and the trace is
\begin{equation}
T_a=\epsilon_{ab}\,b^b\,\delta^2(\mathbf r),\qquad\text{i.e.}\quad T_1=0,\quad T_2=-b\,\delta^2(\mathbf r),
\label{eq:trace}
\end{equation}
with $\epsilon_{12}=1$. The trace points perpendicular to the Burgers vector and, like the torsion, lives at the core. In polar coordinates it has both a radial and an angular component, $T_\rho=T_2\sin\varphi$ and $T_\varphi=T_2\,\rho\cos\varphi$.

\section{Dirac quasiparticles and the valley holonomy}\label{sec:holonomy}

\subsection{Anandan's phase operator for the dislocation}\label{sec:anandan}

For the dislocation~(\ref{eq:metric}) the curvature vanishes and $\omega^{ab}=0$, so only the translational part of Eq.~(\ref{eq:anandan}) survives. On a closed loop $C$ that winds $n_C$ times around the core,
\begin{align}
\Phi_C&=\exp\left(-\frac{i}{\hbar}P_a\oint_C\theta^a\right)=\exp\left(-\frac{i}{\hbar}P_a\int_ST^a\right)\nonumber\\
&=\exp\left(-\frac{i}{\hbar}\,n_C\,\mathbf P\cdot\bb\right),
\label{eq:anandanb}
\end{align}
where we used Eq.~(\ref{eq:torsion}) and $\oint_C dx^a=0$. The phase depends only on the torsion flux through $C$, that is, on the Burgers vector. Neither the shape of $C$ nor the way the torsion is distributed inside the core matters, and a loop that does not enclose the defect gives $\Phi_C=1$. What remains is to identify $\mathbf P$ for the quasiparticles of graphene, which we do in the rest of this section. The answer is $\mathbf P=\hbar(\tau_3\KK+\mathbf q)$, the crystal momentum measured from the $\Gamma$ point. The phase of a loop around a dislocation is then $\exp(-i\tau_3\KK\cdot\bb)\exp(-i\mathbf q\cdot\bb)$: a topological factor that does not depend on energy, and a factor that depends on the envelope momentum $\mathbf q$ and is of order $qa$.

The point we wish to emphasize is that \textit{the Anandan phase of an isolated dislocation never vanishes.} Because $\KK\cdot\bb$ takes only the values $0$ and $\pm2\pi/3$ modulo $2\pi$ (Sec.~\ref{sec:holonomy}B), two cases arise.
\begin{itemize}
\item When $\bb$ lies outside the $\sqrt3\times\sqrt3$\,R$30^\circ$ superlattice, the topological factor is $e^{\mp2\pi i/3}$ in the two valleys. This is always the case for a minimal 5--7 dislocation: all six of its possible Burgers vectors have $\KK\cdot\bb=\pm2\pi/3$. The phase is then independent of the energy, of the shape of the loop, and of the structure of the core.
\item When $\bb$ lies in the superlattice, for instance with $|\bb|=\sqrt3a$ as in the 4--8 dislocation of Sec.~\ref{sec:complexes}, the topological factor equals 1, and the Anandan phase reduces to the dispersive factor $e^{-i\mathbf q\cdot\bb}$.
\end{itemize}
Strictly speaking, the dispersive factor is a translation of the envelope by $\bb$; it reduces to a phase on states of well-defined momentum, such as wave packets, and it grows linearly with energy. Taking $q=E/\hbar v_F$ with $\hbar v_F\approx0.66$~eV\,nm, a carrier at $E=0.1$~eV that goes around a dislocation with $|\bb|=a=0.246$~nm picks up at most $|\mathbf q\cdot\bb|\approx0.04$~rad, and about $0.19$~rad at $0.5$~eV, depending on the angle between $\mathbf q$ and $\bb$. When the topological factor is present it dominates. For dislocations in the superlattice the dispersive factor is the only signature, and it can be recognized by its linear growth with energy and its cosine dependence on the direction of propagation.

\subsection{Translational holonomy}

We use the four-component spinor $\Psi=(\psi_{+A},\psi_{+B},\psi_{-A},\psi_{-B})^T$, where $\pm$ labels the valleys $\pm\KK$, and the representation of Ref.~\cite{mes1},
\begin{equation}
\gamma^0=\tau_1\otimes 1,\quad \gamma^i=-i\tau_2\otimes\sigma_i,\quad \gamma^5=\tau_3\otimes 1,
\label{eq:gammas}
\end{equation}
in which the Pauli matrices $\tau_i$ and $\sigma_i$ act on the valley and sublattice indices. These matrices obey $\{\gamma^a,\gamma^b\}=2\eta^{ab}$ with $\eta=\mathrm{diag}(+,-,-)$, and $\gamma^5$ anticommutes with $\gamma^0$, $\gamma^1$, and $\gamma^2$. In the dislocated space the Dirac equation reads
\begin{equation}
i\gamma^a e_a{}^{\mu}\left(\partial_\mu+\Gamma_\mu\right)\Psi=0 .
\label{eq:dirac}
\end{equation}
With minimal coupling to the full Riemann--Cartan connection, the only part of the torsion that enters the Dirac equation is the axial one~\cite{shap}. Since $S=0$ and $\omega^a{}_b=0$, we have $\Gamma_\mu=0$, and the torsion drops out of the equation altogether. Multiplying by $\gamma^0$ and using $\gamma^0\gamma^i=\tau_3\otimes\sigma_i$, we obtain
\begin{equation}
H=\tau_3\otimes\boldsymbol\sigma\cdot\mathbf p,\qquad p_a=-i\,e_a=-i\,\partial_{X^a},
\label{eq:H}
\end{equation}
which is the free graphene Hamiltonian written in the crystal coordinates $X^a$. In the original coordinates it becomes $-i\,\tau_3\otimes[\sigma_x A^{-1}\partial_x+\sigma_y(\partial_y-BA^{-1}\partial_x)]$. Some earlier treatments write the second term as $-i\chi\Sigma^3(x/r^2)\gamma^1\partial_x$, with $\Sigma^3=1\otimes\sigma_3$; this is the same term, because $\gamma^1\Sigma^3=-i\gamma^2$ in this representation.

The dislocation enters only through the topology of the crystal coordinates: once around the core, $X^a\to X^a+b^a$. On the lattice the wave function is $\Phi(\mathbf X)=\sum_{s=\pm}e^{is\KK\cdot\mathbf X}\psi_s(\mathbf X)$, and it has to be single valued on the lattice sites. Continuing the envelope once around $C$ therefore gives
\begin{equation}
\psi_s(\mathbf X+\bb)=e^{-is\KK\cdot\bb}\,\psi_s(\mathbf X).
\label{eq:twisted}
\end{equation}
Parallel transport around $C$ is the lattice translation by $\bb$, generated by the total crystal momentum $\hat{\mathbf P}=\tau_3\KK+\hat{\mathbf q}$. This is Anandan's operator~(\ref{eq:anandanb}) with $\mathbf P/\hbar=\hat{\mathbf P}$:
\begin{equation}
U(C)=e^{i\bb\cdot\hat{\mathbf P}}=e^{i\tau_3\,\KK\cdot\bb}\;e^{\,\bb\cdot\nabla_X},
\label{eq:holonomy}
\end{equation}
up to the orientation of $C$ (we follow the sign convention of Ref.~\cite{mes1}). The second factor acts on the slowly varying envelope and gives $e^{i\bb\cdot\mathbf q}=1+O(qa)$ near the Dirac points. The first factor is the geometric phase of the dislocation. With $\mathbf a_1=a(1,0)$, $\mathbf a_2=a(1/2,\sqrt3/2)$, and $\KK=(4\pi/3a,0)$, a Burgers vector $\bb=n_1\mathbf a_1+n_2\mathbf a_2$ gives
\begin{equation}
\KK\cdot\bb=\frac{2\pi}{3}\,(2n_1+n_2)\in\left\{0,\pm\frac{2\pi}{3}\right\}\pmod{2\pi},
\label{eq:Kb}
\end{equation}
and $\KK\cdot\bb\equiv0$ exactly when $n_1\equiv n_2\pmod 3$. For the minimal dislocation of Fig.~\ref{fig:lattice}, $\bb=\mathbf a_1$, the two valleys see effective Aharonov--Bohm fluxes of $\mp1/3$ of a flux quantum. Mesaros \textit{et al.} found the same result on the lattice~\cite{mes2,mes1}; here it follows directly from the KV frame. As it must for a defect in a nonmagnetic crystal, the holonomy commutes with the time-reversal operator $\Theta=(\tau_1\otimes\sigma_1)\mathcal K$, which exchanges the valleys.

For a single edge dislocation, $U(C)$ is diagonal in valley space, $U=\mathrm{diag}(e^{i\KK\cdot\bb},e^{-i\KK\cdot\bb})\otimes 1$, so it is an Abelian $U(1)\times U(1)$ phase. It is nonetheless an element of the non-Abelian holonomy group of the honeycomb lattice: once it is combined with the holonomies of disclinations, which rotate the sublattice pseudospin and exchange the valleys~\cite{lamert,mes1}, the elements no longer commute. A wave packet carried adiabatically around the defect~\cite{berry,bohm,mosta,cori} acquires the same $U(C)$, since the transport is a pure translation in the crystal frame. In an experiment, the $1/3$ flux should show up as a shift of one third of a fringe in interference around the dislocation, and as additional wave-front dislocations in the intervalley Friedel oscillations seen by scanning tunneling microscopy, along the lines of Ref.~\cite{dutreix}.

\subsection{The core as a valley flux tube}\label{sec:fluxtube}

Although the torsion is confined to the core, the holonomy~(\ref{eq:holonomy}) is the same for every loop around it. This is how an Aharonov--Bohm flux tube behaves, and the analogy can be made precise. Write $\KK\cdot\mathbf X=\KK\cdot\mathbf x+(\KK\cdot\bb/2\pi)\varphi$ and set $\psi_s=e^{-is\Lambda}\tilde\psi_s$ with $\Lambda=(\KK\cdot\bb/2\pi)\varphi$. The new envelope $\tilde\psi_s$ is single valued and, to leading order in $qa$, obeys
\begin{align}
H&=\tau_3\otimes\boldsymbol\sigma\cdot\left(\mathbf p-\tau_3\,\bm{\mathfrak a}\right),\nonumber\\
\mathfrak a_\mu\,dx^\mu&=K_a\left(\theta^a-dx^a\right)=\frac{\KK\cdot\bb}{2\pi}\,d\varphi .
\label{eq:fluxtube}
\end{align}
The neglected terms, of order $\chi q$, are the envelope translation $e^{\bb\cdot\nabla_X}$ of Eq.~(\ref{eq:holonomy}). The valley gauge field $\bm{\mathfrak a}$ is pure gauge outside the core, and its field strength is the torsion two-form contracted with the crystal momentum:
\begin{equation}
d\mathfrak a=K_a\,T^a=\KK\cdot\bb\;\delta^2(\mathbf r)\,dx\wedge dy .
\label{eq:fluxK}
\end{equation}
The torsion two-form acts as the magnetic field of the translation group, with $\tau_3\KK$ as the charge. Because this charge is odd in the valley index, time reversal is respected. Stokes' theorem gives $\oint_C\bm{\mathfrak a}\cdot d\mathbf l=n_C\,\KK\cdot\bb$ for a loop that winds $n_C$ times around the core. The result is independent of the shape of the loop and of how the torsion is spread over the core, since only the total $\int T^a=b^a$ enters; we checked this explicitly for an asymmetric core~\cite{SM}. The torsion trace $T_\mu$ of Eq.~(\ref{eq:trace}) is another contraction of the same $T^a$, this time with the frame rather than with $\KK$, and it carries no flux (Sec.~\ref{sec:nonminimal}). The global, topological effect of the core is contained in $K_aT^a$, not in $T_\mu$.

This has three consequences.

(i) \textit{$\mathbb Z_3$ topological charge.} The valley flux $\alpha=\KK\cdot\bb/2\pi$ is defined modulo 1 and takes the values $0$, $1/3$, and $2/3$; the fluxes of several enclosed dislocations add. For the Dirac quasiparticles, the topological charge of a dislocation is $\bb$ modulo the sublattice $\{\bb:\KK\cdot\bb\in2\pi\mathbb Z\}$. Section~\ref{sec:complexes} extends this to arbitrary defect complexes.

(ii) \textit{Robustness, and what the core does.} Short-range terms at the core, such as the axial contact term of Sec.~\ref{sec:nonminimal} or scalar and intervalley terms, cannot change $\alpha$; they enter only through the boundary condition at the flux line. For massless Dirac fermions in a flux $0<\alpha<1$, every angular-momentum channel $j=m+\frac12$ except $m=0$ has a solution that is regular in both spinor components. In the $m=0$ channel one component is necessarily singular, although square integrable, and a one-parameter family of boundary conditions (self-adjoint extensions) is allowed~\cite{gerbert,hagen}. We verified this channel structure explicitly~\cite{SM}. The core thus plays two separate roles: its torsion flux fixes the Aharonov--Bohm phase shifts in all channels, while its microscopic structure selects the boundary condition in the single channel $m=0$.

(iii) \textit{Observable consequences.} At low energy, carriers scatter off an isolated dislocation as off an Aharonov--Bohm flux of $\pm1/3$ in the two valleys. The cross section is proportional to $\sin^2(\pi\alpha)/k=3/(4k)$, up to the dependence on the $m=0$ boundary condition, and it exceeds the contact contribution computed in Sec.~\ref{sec:nonminimal}.

\section{Loops enclosing defect complexes}\label{sec:complexes}

\begin{figure*}[t]
\includegraphics[width=0.82\textwidth]{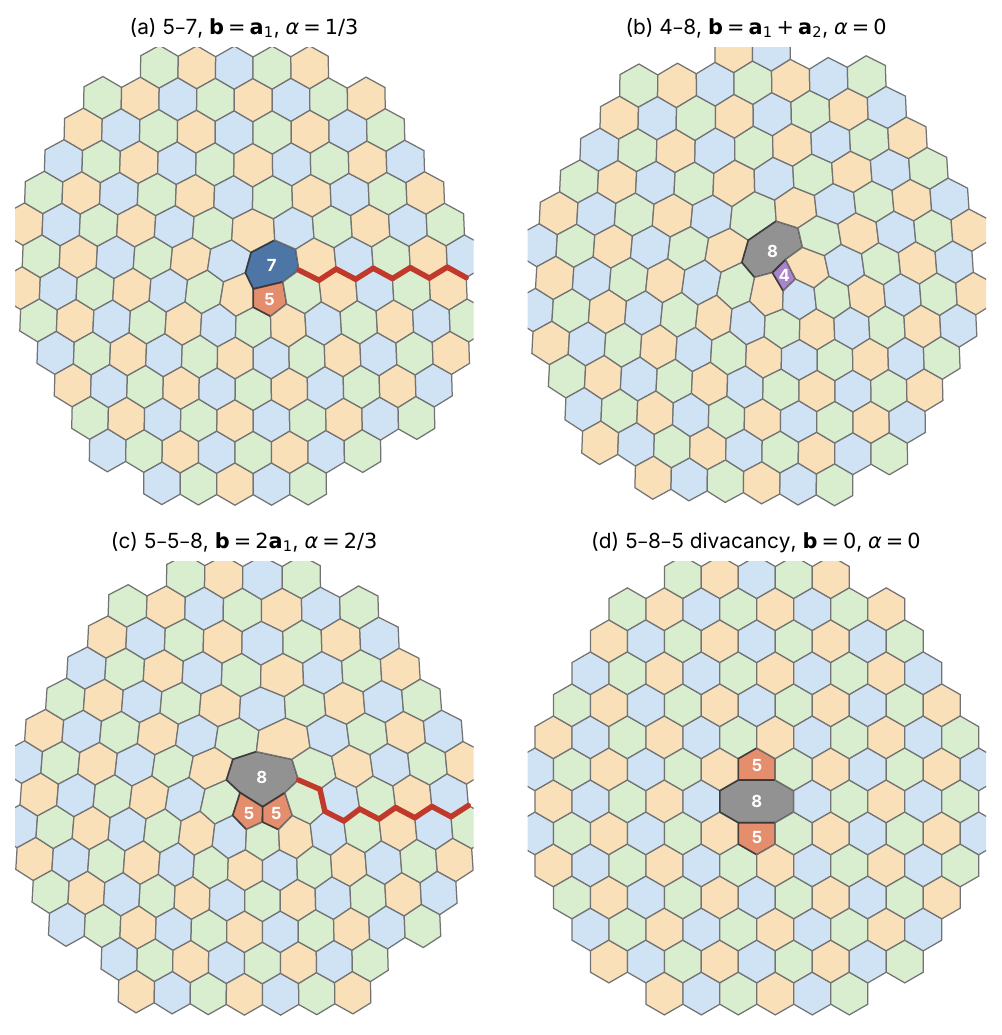}
\caption{Kekul\'e three-coloring of the hexagons around four relaxed atomic models; non-hexagonal rings are labeled by their size. Colors are propagated from hexagon to hexagon with $c(\mathbf x+\mathbf t)=c(\mathbf x)+3\KK\cdot\mathbf t/2\pi \pmod 3$. When the valley flux $\alpha=\KK\cdot\bb/2\pi$ is nonzero the coloring cannot be closed around the core, and a seam (red) runs from the core to the boundary: (a) 5--7 dislocation with $\bb=\mathbf a_1$; (c) 5--5--8 dislocation with $\bb=2\mathbf a_1$. There is no seam for (b) the 4--8 dislocation with $\bb=\mathbf a_1+\mathbf a_2$, nor for (d) the 5--8--5 divacancy, even though (c) and (d) have the same rings.}
\label{fig:kekule}
\end{figure*}

\subsection{Volterra elements and the holonomy of a complex}

A ring with $n$ sides centered at the hexagon site $\mathbf x$ is a wedge disclination with Frank angle $\Omega=(6-n)\pi/3$. Placing the origin at a hexagon center makes $\mathbf x$ a lattice vector. The Volterra element of the ring is the Euclidean motion $g=(R_\Omega,(1-R_\Omega)\mathbf x)$, which acts as $\mathbf r\to R\mathbf r+\mathbf t$~\cite{kleinert,katrev}. A loop $C$ that encloses rings $1,\dots,N$, taken counterclockwise, has holonomy $g_1g_2\cdots g_N$, with the product rule $(R_1,\mathbf t_1)(R_2,\mathbf t_2)=(R_1R_2,\mathbf t_1+R_1\mathbf t_2)$. This is just the path-ordered Anandan operator~(\ref{eq:anandan}): its rotational part records the Frank angles and its translational part records the torsion. In a flat sheet the Frank angles of a complex must add up to zero (otherwise the sheet turns into a cone), and the product reduces to a translation by the net Burgers vector
\begin{equation}
\bb=\sum_{i=1}^{N}R_1\cdots R_{i-1}\left(1-R_i\right)\mathbf x_i .
\label{eq:bnet}
\end{equation}
This vector does not depend on where the origin is placed~\cite{SM}.

On the low-energy states, a lattice translation $\mathbf t$ acts as $e^{i\tau_3\KK\cdot\mathbf t}$ (Sec.~\ref{sec:holonomy}). A rotation by $\pi/3$ about a hexagon center swaps the sublattices and sends $\KK$ to $-\KK$ up to a reciprocal vector $\mathbf G$, so it exchanges the valleys; a rotation by $2\pi/3$ leaves both unchanged~\cite{lamert,mes1}. A loop that encloses an odd number of pentagons and heptagons has a non-Abelian, valley-exchanging holonomy~\cite{lamert,mes1}, but such a loop has to cut through a complex whose net Frank angle is zero. For loops that enclose whole complexes, the rotations multiply to the identity and
\begin{equation}
U(C)=\exp\left(i\tau_3\,\KK\cdot\bb\right),
\label{eq:Ucomplex}
\end{equation}
with $\bb$ from Eq.~(\ref{eq:bnet}). Every loop around a complex carries the valley flux $\alpha=\KK\cdot\bb/2\pi$ modulo 1, and the Berry phase of the complex is set by its net Burgers vector.

\subsection{Three-coloring rule}

Using $\KK\cdot(M\mathbf v)=(M^{-1}\KK)\cdot\mathbf v$ for rotations, together with $R_{2\pi/3}^{-1}\KK\equiv\KK$ and $R_{\pi/3}^{-1}\KK\equiv-\KK$ modulo $\mathbf G$ and $\mathbf G\cdot\mathbf x\in2\pi\mathbb Z$, Eq.~(\ref{eq:bnet}) becomes
\begin{equation}
\KK\cdot\bb\equiv-\sum_{i\,\in\,\{5,7\}}\epsilon_i\,\KK\cdot\mathbf x_i\pmod{2\pi}.
\label{eq:colour}
\end{equation}
The sum runs only over pentagons and heptagons, and $\epsilon_i=(-1)^{p_i}$, where $p_i$ is the number of pentagons and heptagons that come before $i$ along the loop. The quantity $c(\mathbf x)=3\KK\cdot\mathbf x/2\pi \bmod 3$ is the color of the hexagon site in the Kekul\'e ($\sqrt3\times\sqrt3$\,R$30^\circ$) three-coloring of the honeycomb. Three consequences follow; we checked all of them on 3000 random complexes using exact lattice arithmetic~\cite{SM}.

(i) \textit{Squares and octagons never contribute.} A complex made only of 4- and 8-membered rings has $\alpha=0$, whatever its Burgers vector. That vector necessarily lies in the $\sqrt3\times\sqrt3$ superlattice, where $\KK\cdot\bb\in2\pi\mathbb Z$.

(ii) \textit{Pentagon--heptagon pairs.} For a 5--7 pair, $\alpha=\mp(c_5-c_7)/3$. The pair carries a flux of $\pm1/3$ when the two rings sit on hexagon sites of different color, which is always true for adjacent rings. It carries no flux when they sit on sites of the same color, as second neighbors do, even though $|\bb|=\sqrt3a\neq0$.

(iii) \textit{$\mathbb Z_3$ addition.} The fluxes of separate complexes add modulo 1. Stone--Wales defects and divacancies have $\bb=0$, and pairs or triples of dislocations whose Burgers vectors add up to a superlattice vector cannot be seen from far away.

\subsection{Atomic models}

\begin{table*}[t]
\caption{Atomic models of defects in graphene, and the Anandan phase of a loop around each of them. The core rings are those within $4a$ of the defect. The Burgers vector $\bb$ was measured with Burgers circuits through hexagons at radii $5.5a$, $7.5a$, and $9.5a$, which all give the same answer; no hop deviates from an ideal lattice vector by more than $0.09a$. The valley flux $\alpha=\KK\cdot\bb/2\pi$ is given modulo 1 and up to the sign that distinguishes the two valleys. The last column lists $\Phi_C=\exp(-i\tau_3\KK\cdot\bb)\exp(-i\mathbf q\cdot\bb)$ for the valley $+\KK$; for $-\KK$ the topological factor is complex conjugated.}
\label{tab:models}
\begin{ruledtabular}
\begin{tabular}{llccl}
Model & Core rings & $\bb$ & $|\alpha|$ & Anandan phase $\Phi_C$ \\ \hline
5--7 dislocation & 5, 7 & $\mathbf a_1$ & $1/3$ & $e^{+2\pi i/3}\,e^{-i\mathbf q\cdot\bb}$ (topological and dispersive)\\
5--7 dislocation (second neighbors) & 5, 7 & $\mathbf a_1+\mathbf a_2$ & $0$ & $e^{-i\mathbf q\cdot\bb}$ (dispersive only)\\
4--8 dislocation & 4, 8 & $\mathbf a_1+\mathbf a_2$ & $0$ & $e^{-i\mathbf q\cdot\bb}$ (dispersive only)\\
5--5--8 dislocation & 5, 5, 8 & $\mathbf a_1+\mathbf a_2$ & $0$ & $e^{-i\mathbf q\cdot\bb}$ (dispersive only)\\
5--5--8 dislocation & 5, 5, 8 & $2\mathbf a_1$ & $1/3$ & $e^{-2\pi i/3}\,e^{-i\mathbf q\cdot\bb}$ (topological and dispersive)\\
Stone--Wales & 5, 5, 7, 7 & $0$ & $0$ & $1$\\
divacancy & 5, 5, 8 & $0$ & $0$ & $1$\\
\end{tabular}
\end{ruledtabular}
\end{table*}

To see whether Eqs.~(\ref{eq:Ucomplex}) and (\ref{eq:colour}) hold for realistic structures, we built relaxed atomic models of seven defects~\cite{SM}. The dislocations come from the Volterra construction: we removed a strip of width $|\bb|$ and relaxed the lattice with the isotropic elastic field of an edge dislocation, using a Poisson ratio of $0.17$. Placing the cut at different positions produces different reconstructions of the core for the same $\bb$. The Stone--Wales defect was made by rotating one bond by $90^\circ$, and the divacancy by removing a dimer. In every case we rebuilt the network from nearest neighbors and paired the under-coordinated atoms at the core. We then identified the rings and measured $\bb$ with Burgers circuits that run through hexagons only and never touch the core. Table~\ref{tab:models} collects the results.

All seven models agree with the coloring rule, and two points deserve comment. The first is that the rings do not determine the Berry phase. A 5--5--8 core can be a point defect ($\bb=0$, the divacancy), a dislocation without valley flux ($\bb=\mathbf a_1+\mathbf a_2$), or a dislocation with flux $\pm1/3$ ($\bb=2\mathbf a_1$). The second is that the 4--8 dislocation, built only from even-membered rings, has $|\bb|=\sqrt3a$ and yet no valley flux, as rule (i) demands; the same is true of a 5--7 pair on second-neighbor sites. Among the structures usually observed, the ones that are topologically visible to the quasiparticles are the dislocations whose Burgers vector lies outside the $\sqrt3\times\sqrt3$ superlattice, the minimal 5--7 pair being the most important. The last column of Table~\ref{tab:models} shows, however, that every dislocation still has a nonzero Anandan phase, because the dispersive factor survives whenever $\bb\neq0$. Only defects with $\bb=0$, such as Stone--Wales defects and divacancies, have a trivial phase on every loop around them.

\subsection{Kekul\'e coloring as a direct image of the flux}

Equation~(\ref{eq:colour}) has a simple geometric meaning. Color the hexagons with $c(\mathbf x)$, passing from one hexagon to the next with $c(\mathbf x+\mathbf t)=c(\mathbf x)+3\KK\cdot\mathbf t/2\pi$. Around a complex with $\alpha\neq0$ the coloring cannot be closed, so a seam has to run from the core to the boundary, and across it the color jumps by $3\alpha$ modulo 3 (Fig.~\ref{fig:kekule}). The color encodes the phase $\KK\cdot\mathbf x$ of the intervalley interference pattern that scanning tunneling microscopy resolves as a $\sqrt3\times\sqrt3$\,R$30^\circ$ modulation around impurities~\cite{dutreix}. A seam is a $2\pi/3$ phase slip of this pattern, i.e., an extra wave-front dislocation, which suggests that the valley flux of a defect could be read off directly from a topographic image.

When many defects are present, a loop picks up the sum of the valley fluxes of the complexes it encloses, modulo 1. To the Dirac quasiparticles, polycrystalline graphene then looks like a random array of $\mathbb Z_3$ valley flux tubes sitting at the dislocations, on top of short-range scatterers (point defects and the cores themselves) that carry no flux. A grain boundary made of 5--7 pairs with a common Burgers vector contributes $N\KK\cdot\bb$ to a loop that encloses $N$ of its pairs.

\section{Nonminimal coupling to the torsion trace}\label{sec:nonminimal}

\subsection{Time-reversal symmetry selects the axial coupling}

Since the axial torsion vanishes, the only way the torsion trace~(\ref{eq:trace}) can act on the quasiparticles is through a nonminimal term~\cite{shap}. At linear order there are two Lorentz-covariant possibilities,
\begin{equation}
\mathcal L_{\rm nm}=\nu\,\bar\Psi\gamma^\mu T_\mu\Psi+\nu_5\,\bar\Psi\gamma^5\gamma^\mu T_\mu\Psi .
\label{eq:Lnm}
\end{equation}
With $\gamma^0\gamma^5\gamma^i=-1\otimes\sigma_i$, their contribution to the Hamiltonian is
\begin{equation}
H_{\rm nm}=-\nu\,\tau_3\otimes\boldsymbol\sigma\cdot\mathbf T+\nu_5\,1\otimes\boldsymbol\sigma\cdot\mathbf T ,
\label{eq:Hnm}
\end{equation}
with $\mathbf T=(T_1,T_2)$. The first term has exactly the form of an electromagnetic vector potential, $H=\tau_3\otimes\boldsymbol\sigma\cdot(\mathbf p+e\mathbf A)$. Consider the physical time-reversal operation $\Theta=(\tau_1\otimes\sigma_1)\mathcal K$, which leaves $H_0$ invariant, exchanges the valleys, and satisfies $\Theta^2=+1$. For a real, static $\mathbf T$ one finds $\Theta(\tau_3\otimes\boldsymbol\sigma\cdot\mathbf T)\Theta^{-1}=-\tau_3\otimes\boldsymbol\sigma\cdot\mathbf T$, whereas $\Theta(1\otimes\boldsymbol\sigma\cdot\mathbf T)\Theta^{-1}=+1\otimes\boldsymbol\sigma\cdot\mathbf T$. The conclusion does not depend on which valley-exchanging antiunitary symmetry of $H_0$ with $\Theta^2=+1$ one takes as the physical one~\cite{SM}. Torsion is a property of the static lattice and is even under time reversal, so a nonmagnetic crystal cannot produce the vector coupling, and $\nu=0$ in graphene. What remains is the axial coupling, which enters each valley with the opposite sign relative to the momentum, exactly as the strain-induced pseudo-gauge field does~\cite{vozrep,jackiw,cortijo2015}. The situation is different for a fundamental Dirac fermion in $3+1$ dimensions. There the time-reversal operator does not mix internal indices, and for a static torsion both $\nu\bar\Psi\gamma^\mu T_\mu\Psi$ and $\nu_5\bar\Psi\gamma^5\gamma^\mu T_\mu\Psi$ are odd~\cite{SM}. In graphene the four spinor components carry sublattice and valley indices rather than spin, and the axial coupling survives only because the physical time reversal exchanges the valleys. Nonminimal couplings designed for fundamental fermions cannot simply be carried over to graphene without checking them against the symmetries of the lattice. Inserting Eq.~(\ref{eq:trace}),
\begin{equation}
H_5=-\hbar v_F\,\nu_5\,1\otimes\boldsymbol\sigma\cdot(\hat{\mathbf z}\times\bb)\,\delta^2(\mathbf r)\equiv\hbar v_F\,\lambda\,1\otimes\sigma_y\,\delta^2(\mathbf r),
\label{eq:H5}
\end{equation}
where $\lambda=-\nu_5 b$ for $\bb=b\hat{\mathbf x}$ has the dimension of length.

\subsection{Absence of a geometric phase for loops enclosing the defect}

Equation~(\ref{eq:H5}) is a valley-odd vector potential $\tau_3\boldsymbol{\mathcal A}$ with $\boldsymbol{\mathcal A}=\nu_5\mathbf T$. Locally it can be removed with the Dirac phase factor $\Psi=\exp(-i\tau_3\int^{\mathbf r}\boldsymbol{\mathcal A}\cdot d\mathbf l)\Psi_0$~\cite{dirac}, so any geometric phase would have to come from the circulation $\oint_C\boldsymbol{\mathcal A}\cdot d\mathbf l$. For a general distribution of dislocations with Burgers-vector density $\alpha^a(\mathbf r)$ one has $T_a=\epsilon_{ab}\alpha^b$, and Stokes' theorem gives
\begin{equation}
\oint_{\partial S}T_a\,dX^a=\int_S\left(\partial_1T_2-\partial_2T_1\right)d^2X=-\oint_{\partial S}\bm\alpha\cdot\hat{\mathbf n}\,dl .
\label{eq:theorem}
\end{equation}
The circulation of the torsion trace thus measures the flux of Burgers-vector density through the loop itself, not the Burgers vector that the loop encloses. For a dislocation localized at the origin it vanishes on every loop that goes around the defect without crossing the core, and these are exactly the loops of interest here. The corresponding pseudo-magnetic field, $\partial_1\mathcal A_2-\partial_2\mathcal A_1=-\nu_5 b\,\partial_x\delta^2(\mathbf r)$, is a point dipole with no net flux. If the core is smeared over a width $a$, the circulation is nonzero only for loops whose boundary passes through the core, and then it depends on the path (Ref.~\cite{SM} gives an example), so it is not a topological phase. The same is true of the Mead--Berry connection of a wave packet $\psi(\mathbf r-\mathbf R)$ moved along $\mathbf R(t)$: the contribution of $H_5$ is $\nu_5\int|\psi(\mathbf r-\mathbf R)|^2\mathbf T\,d^2r$, and its loop integral vanishes as long as the packet stays clear of the core.

Whether it is coupled vectorially or axially, then, the torsion trace does not produce an Abelian phase proportional to $\bb$. A phase of the form $\Phi=2\pi\nu\chi=\nu b$ cannot be correct in any case. With $\nu$ dimensionless it has the dimension of length, and this remains true in natural units, $\hbar=v_F=1$, which still leave length as an independent dimension. It could be made dimensionless only by giving $\nu$ the dimension of an inverse length, so that it plays the role of a wave number, or by measuring lengths in units of the lattice constant, which merely hides the scale. The first option is instructive: the dimensionally consistent product of a wave number and the Burgers vector is precisely Anandan's phase $\mathbf P\cdot\bb/\hbar$ of Eq.~(\ref{eq:anandanb}), in which the wave number is the crystal momentum $\tau_3\KK+\mathbf q$ rather than a free coupling constant. Independently of units, the circulation~(\ref{eq:theorem}) vanishes for every loop that encloses the defect. The topological information about the Burgers vector is carried by the holonomy~(\ref{eq:holonomy}), not by the trace.

The long-range pseudo-gauge field produced by the strain of the dislocation behaves in the same way. It is $\boldsymbol{\mathcal A}_{\rm s}\propto(\beta/a_{cc})(u_{xx}-u_{yy},-2u_{xy})$, with $\beta=-\partial\ln t/\partial\ln a_{cc}\approx2$--$3$~\cite{vozrep}. For the displacement field of Eq.~(\ref{eq:metric}), and also for the full isotropic elastic solution at any Poisson ratio, its circulation on every circle around the core vanishes, for both lattice orientations~\cite{SM}. At linear order, the pseudo-magnetic field of a dislocation has no monopole moment.

\subsection{Short-range scattering}

What $H_5$ does produce is scattering. Take a conduction-band state $|\mathbf k\rangle$ incident at angle $\theta$, measured from $\bb$, and scattered to angle $\theta'$. The matrix element is $\langle\mathbf k'|\sigma_y|\mathbf k\rangle$, and $|\langle\mathbf k'|\sigma_y|\mathbf k\rangle|^2=\sin^2[(\theta+\theta')/2]$ in both valleys. In the Born approximation the differential cross section is
\begin{equation}
\frac{d\sigma}{d\theta'}=\frac{\lambda^2k}{2\pi}\,\sin^2\frac{\theta+\theta'}{2},
\label{eq:dsigma}
\end{equation}
which gives the total and transport cross sections
\begin{equation}
\sigma=\frac{\lambda^2k}{2},\qquad \sigma_{\rm tr}(\theta)=\lambda^2k\left(\frac12+\frac{\cos2\theta}{4}\right).
\label{eq:sigmas}
\end{equation}
Two features set this channel apart from an ordinary short-range scalar impurity. Backscattering is allowed: for $\theta=0$ and $\theta'=\pi$ the matrix element has unit modulus, whereas a scalar potential gives zero~\cite{castro}. And the transport cross section is anisotropic, three times larger for carriers moving parallel to $\bb$ than for carriers moving perpendicular to it. Equations~(\ref{eq:dsigma}) and (\ref{eq:sigmas}) hold to leading order in $k\lambda\ll1$; beyond that order the contact interaction must be regularized at the size of the core.

To estimate $\nu_5$, note that a 5--7 core changes the hopping on a few bonds by some amount $\delta t$. Projected onto the Dirac envelopes, the intravalley part of a bond perturbation is purely off-diagonal in the sublattice index, with strength $\delta t\,A_c\,\delta^2(\mathbf r)$, where $A_c=3\sqrt3a_{cc}^2/2$ is the area of the unit cell. Comparing with Eq.~(\ref{eq:H5}) and using $\hbar v_F=3ta_{cc}/2$ gives $|\lambda|\sim a\,|\delta t|/t$ with $a=\sqrt3a_{cc}$, and hence $|\nu_5|\sim|\delta t|/t$, of order $0.1$--$1$ for $|\bb|=a$. A quantitative value, along with the scalar and intervalley contact terms that the core also generates, would require a microscopic calculation of the core~\cite{tor}. The contact cross section grows as $k$, while Aharonov--Bohm scattering by the valley flux~(\ref{eq:Kb}) goes as $1/k$~\cite{ab}. At low carrier density the topological holonomy dominates, and the torsion-trace term is a small short-range correction.

\section{Conclusions}\label{sec:conclusions}

We have studied the Berry phase of graphene quasiparticles on loops that enclose defects, and only on such loops. The tools were Anandan's generalization of the geometric phase to spaces with torsion, i.e., the Poincar\'e holonomy~(\ref{eq:anandan}), the Katanaev--Volovich geometry, and the Volterra composition of disclinations. In this setting the torsion enclosed by a loop couples to the momentum, and for graphene that momentum is the crystal momentum $\tau_3\KK$.

The main conclusion is that the Anandan phase of an isolated dislocation never vanishes. On any loop around it the phase is $\exp(-i\tau_3\KK\cdot\bb)\exp(-i\mathbf q\cdot\bb)$. The topological factor $e^{\mp2\pi i/3}$ is present whenever $\bb$ lies outside the $\sqrt3\times\sqrt3$ superlattice, which includes every minimal 5--7 dislocation, and it reproduces the lattice result of Mesaros \textit{et al.}, now understood as a torsion-flux phase. The dispersive factor is present for every $\bb\neq0$, so dislocations in the superlattice, such as 4--8 cores, still carry a purely dispersive Anandan phase. Although the torsion is confined to the core, it acts on every loop that encloses it: the field strength of the valley gauge field is $K_aT^a$, and the core behaves as an Aharonov--Bohm flux tube of the translation group.

For a general complex with zero net Frank angle the holonomy is $\exp(i\tau_3\KK\cdot\bb_{\rm net})$, and the valley flux follows a three-coloring rule. Only pentagons and heptagons contribute, through the Kekul\'e color of the sites they occupy; squares and octagons never do. Atomic models of seven common defects follow the rule and show that the rings alone do not determine the phase: the same 5--8--5 core can be invisible or carry a flux of $\pm1/3$, and a 4--8 dislocation carries none.

The flux of a defect can be read from the seam it forces in the Kekul\'e coloring, which tracks the phase of the intervalley pattern seen in scanning tunneling microscopy. The microscopic structure of the core enters only through the boundary condition in a single angular-momentum channel.

The torsion trace, on the other hand, carries no flux. A vector coupling to it would break time-reversal symmetry, and the axial coupling that symmetry allows leads only to short-range, anisotropic scattering in which backscattering is allowed. Phase-sensitive experiments should look for the $\mathbb Z_3$ fluxes of dislocations, which are a topological property of the defect, rather than for a phase that varies continuously with a coupling constant.

\begin{acknowledgments}
We thank K. Bakke for valuable discussions. This work was partially supported by CNPq and CAPES. C.F. acknowledges support from CNPq through a Research Productivity Fellowship (Grant No.~311781/2021-7). G.Q.G. thanks FAPESQ-PB for financial support (Grant No.~BLD-ADT-A2377/2024).
\end{acknowledgments}

\section*{Declaration of AI use}

During the preparation of this work the authors used Claude
(Anthropic) to check the analytical derivations, to
generate the figures, and to improve the language of the manuscript. After using this tool, the authors reviewed and edited the content as needed and take full responsibility for the content of the publication.

\end{document}